\documentclass[twocolumn,english]{svjour3}
\usepackage{newcent}
\usepackage[T1]{fontenc}
\usepackage[latin9]{inputenc}
\usepackage{units}
\usepackage{url}
\usepackage{amstext}
\usepackage{graphicx}

\makeatletter

\providecommand{\tabularnewline}{\\}

\makeatother

\usepackage{babel}
\begin{document}

\title{Velocity measurements in the liquid metal flow driven by a two-phase
inductor}

\author{A. Pedcenko \and A. Bojarevi\v{c}s \and\\
J. Priede \and G. Gerbeth \and R. Hermann}

\institute{Coventry University, Applied Mathematics Research Centre, Priory
Street, Coventry, CV1 5FB, UK \and Institute of Physics, University
of Latvia, Salaspils, LV-2169, Latvia \and Coventry University, Applied
Mathematics Research Centre, Priory Street, Coventry, CV1 5FB, UK
\and Helmholtz-Zentrum Dresden-Rossendorf, P.O. Box 510119, D-01314
Dresden, Germany \and Leibniz Institute for Solid State and Materials
Research (IFW) Dresden, Germany}
\maketitle
\begin{abstract}
We present the results of velocity measurements obtained by ultrasonic
Doppler velocimetry and local potential probes in the flow of GaInSn
eutectic melt driven by a two-phase inductor in a cylindrical container.
This type of flow is expected in a recent modification to the floating
zone technique for the growth of small-diameter single intermetallic
compound crystals. We show that the flow structure can be changed
from the typical two toroidal vortices to a single vortex by increasing
the phase shift between the currents in the two coils from $0^{\circ}$
to $90^{\circ}$ degrees. The latter configuration is thought to be
favourable for the growth of single crystals. The flow is also computed
numerically, and a reasonable agreement with the experimental results
is found. The obtained results may be useful for the design of combined
two-phase electromagnetic stirrers and induction heaters for metal
or semiconductor melts. 
\end{abstract}

\section{Introduction}

The present work is concerned with a physical modelling of the melt
flow in a recent modification to the floating zone technique for the
growth of small-diameter single intermetallic compound crystals \cite{Patent}.
The quality of crystals grown by this method depends on the growth
conditions, particularly on the shape of the solidification front.
The growth of single crystals usually requires a convex solidification
front \cite{Rudolph-10}. The shape of this front can strongly be
affected by the convective heat transport in the melt. A standard
single-phase induction heater usually drives a radially inward jet
of hot melt at the middle of the floating zone \cite{Hermann-etal-01}.
At the centre of the floating zone, the radial jet splits into two
nearly symmetric axial jets, which further stream towards both solid-liquid
interfaces. The hot axial jet impinging at the centre of the growth
interface renders it concave so promoting a polycrystalline growth.
To overcome this adverse effect, we recently proposed a two-phase
inductor, which is able to reverse the direction of the electromagnetically
driven melt flow so that the solidification front becomes mostly convex
\cite{Herm-etal05}.

The inductor consists of two coils that are fed by alternating currents
of the same amplitude but a 90 degrees phase shift between them. In
the original design for radio-frequency induction heating, the phase
lag is achieved by short circuiting the secondary coil through a variable
capacitor that is tuned so that the resonance frequency of the secondary
circuit coincides with the frequency of the power supply \cite{PG}.
In this way, a component of the magnetic field is created that travels
towards the coil with the phase lag so dragging the liquid metal along
it. As a result, the outer layers of the hot melt are driven towards
the periphery of the solidification front while the cooled down melt
returns in the centre \cite{Rudolph-08}. A similar multiphase stirrer
for low conductivity melts has been developed by Ernst \emph{et al.
}\cite{Ernst-etal-05a}, who also used it for refining the grain size
in gold alloys \cite{Ernst-etal-05b}.

Until now, there have been no direct experimental observations of
the effect of the phase shift in a two-coil inductor on the structure
of ensuing fluid flow. In this paper, we present comprehensive results
of velocity measurements in a model experiment using GaInSn eutectic
alloy driven by a simple two-phase stirrer in a cylindrical container.
The evolution of the flow pattern has been determined depending on
the phase shift between the two coils which were connected to separate
power supplies with externally controllable phase shifts. The experiments
were also modelled numerically and numerical results compared with
the measurements.

The paper is organized as follows. In the following two sections,
experimental setup and measurement techniques are described. Experimental
results are presented in Sec. \ref{sec:expres} and compared with
numerical results in Sec. \ref{sec:comp}. The paper is concluded
by a summary in Sec. \ref{sec:sum}.

\section{\label{sec:setup}Experimental setup}

\begin{figure*}
\begin{centering}
\includegraphics[height=0.28\textheight]{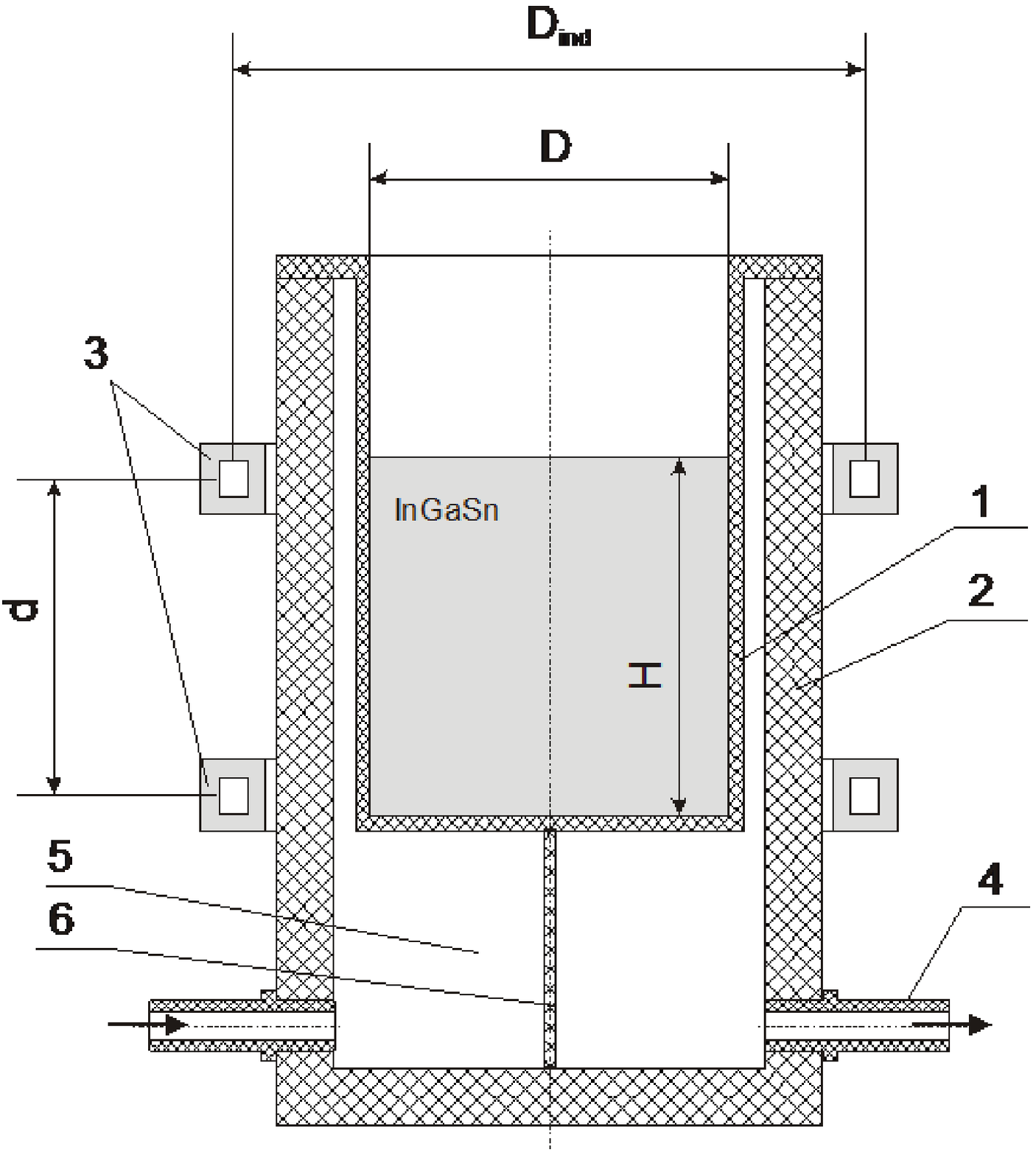}\includegraphics[height=0.28\textheight]{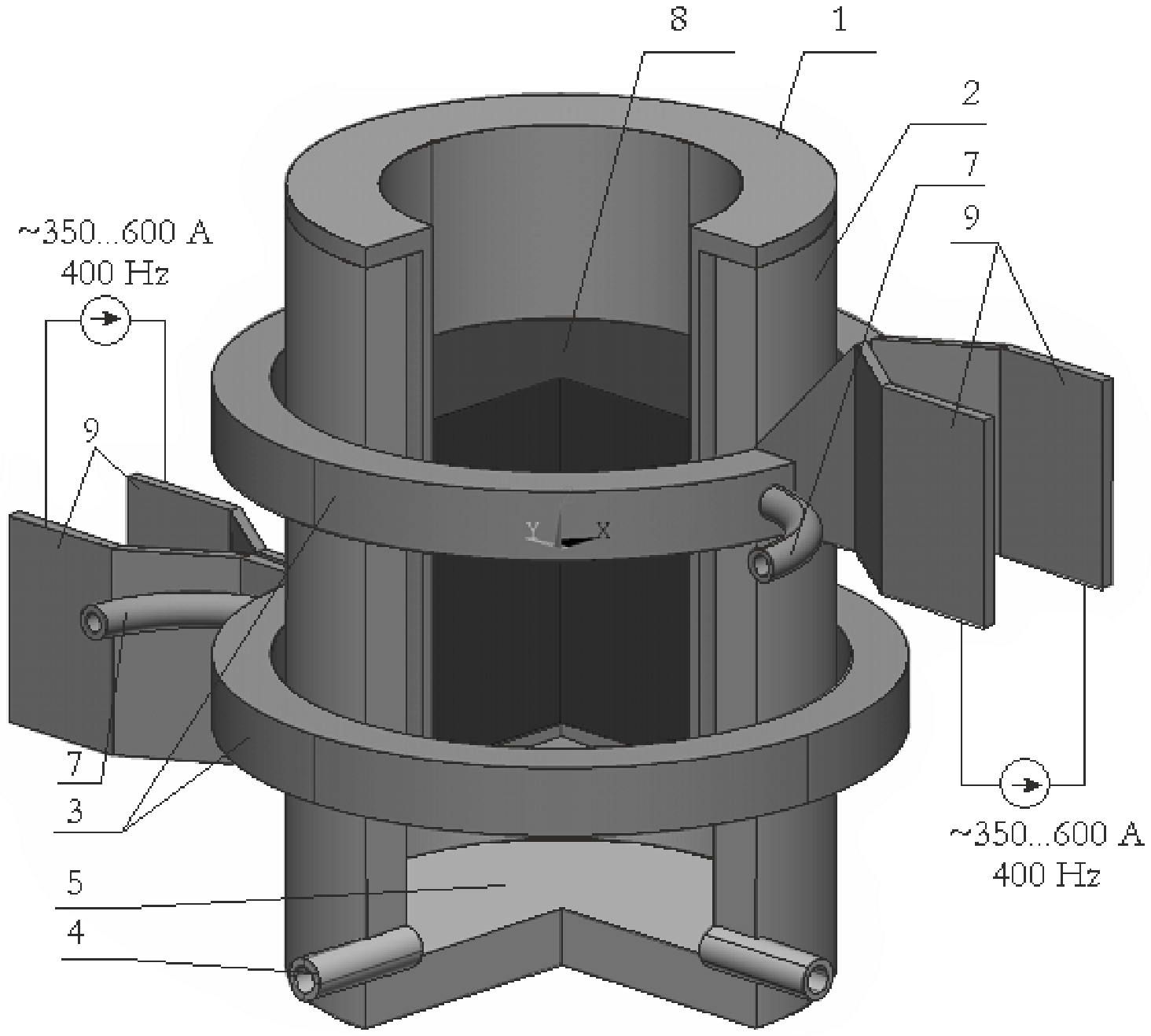}\\
 (a)\hspace{0.5\columnwidth}(b) 
\par\end{centering}

\caption{\label{fig:setup}Experimental setup in cross-section (a) and 3D (b)
views showing the inner container of diameter $D=\unit[50]{mm}$ with
liquid metal (1), the outer container with cooling water (2), two
single-winding coils of diameter $D_{\mathrm{ind}}=\unit[88]{mm}$
and vertical separation $d=\unit[44]{mm}$ (3), one of the six cooling
water intakes and outlets distributed uniformly along the circumference
(4), water jacket (5), cooling water partitioning walls (6), water
intakes and outlets for coil cooling (7), GaInSn alloy (8), coil terminals
(9). }
\end{figure*}

The main element of the experimental setup shown in Fig. \ref{fig:setup}
is an open top cylindrical container of $\unit[50]{mm}$ in both the
diameter and height (unity aspect ratio) made of a $\unit[2]{mm}$-thick
Polycarbonate. The container was filled with GaInSn alloy at eutectic
composition with $67\,\%$ Ga, $20.5\,\%$ In and $12.5\,\%$ Sn,
and kept in a water jacket at $\unit[15]{\mathrm{^{\circ}C}}$ temperature.
The relevant physical properties of GaInSn are listed in Table \ref{tab:param}.
The top surface of the liquid metal was kept open for measurements.
The Joule heat generated in the liquid metal was efficiently removed
by passing the cooling water through a labyrinth of thin partitioning
walls, which were uniformly distributed along the side wall of the
container. The container was placed inside an inductor consisting
of two circular copper coils of diameter $D_{\mathrm{ind}}=\unit[88]{mm},$
axial separation $d=\frac{1}{2}D_{\mathrm{ind}}$ and a rectangular
$\unit[8\times9]{mm^{2}}$ cross-section with an inner cooling channel.
To minimize the perturbation of the magnetic field by the coil terminals,
the latter were made of a $\unit[2]{mm}$-thick copper strip welded
into each copper winding with a $\unit[1]{mm}$ gap, which was filled
with Teflon. Thus the current in each coil was interrupted only over
this small gap which constituted less than $0.4\,\%$ of the total
coil circumference. To minimize the superposition of perturbations,
the terminals of each coil were placed across the diameter of the
container as shown in Fig. \ref{fig:setup}b.

\begin{table}
\begin{centering}
\begin{tabular}{|c|c|}
\hline 
density  & $\rho=\unit[6360]{kg/m^{3}}$\tabularnewline
\hline 
electrical conductivity & $\sigma=\unit[3.2\times10^{6}]{\unit{S/m}}$\tabularnewline
\hline 
kinematic viscosity  & $\nu=\unit[3.4\times10^{-7}]{\unit{m^{2}/s}}$\tabularnewline
\hline 
container diameter $=$ height  & $H=D=\unit[0.05]{m}$\tabularnewline
\hline 
diameter of coils  & $D_{\textit{ind}}=\unit[0.088]{m}$\tabularnewline
\hline 
distance between coils  & $d=\unit[0.044]{m}$\tabularnewline
\hline 
current amplitude  & $I_{0}=100\cdots\unit[600]{A}$\tabularnewline
\hline 
AC frequency  & $f=\unit[400]{Hz}$\tabularnewline
\hline 
\end{tabular}
\par\end{centering}

\caption{\label{tab:param}Physical properties of the GaInSn alloy and experimental
parameters.}
\end{table}

The coils were powered as follows. First, two source signals with
preset amplitudes, frequencies and relative phases were generated
digitally by a computer program and then converted into analogue form
by a D/A converter. Second, the analogue signals were amplified by
a two-channel power amplifier. Two toroidal transformers were used
to match the output impedance of the amplifier with that of the inductor.
This allowed us to reach the RMS current up to $\unit[600]{A}$ in
each coil, which produced magnetic field with the amplitude of vertical
component $B_{0}=\unit[17]{mT}$ in the centre of inductor. As the
phase shift between the currents in both coils was increased from
$\Delta\phi=0^{\circ}$ to $90^{\circ},\,120^{\circ},\,150^{\circ}$
the amplitude decreased respectively to $B_{0}\approx12,\,8,\,\unit[4.5]{mT}.$
At $\Delta\phi=180^{\circ}$ the vertical components of the magnetic
field produced by the opposite currents in both coils nearly compensated
each other in the mid-plane of the inductor. The magnetic field in
the liquid metal was further reduced by the electromagnetic skin effect
with the characteristic thickness $\delta=1/\sqrt{\mu_{0}\sigma\omega/2},$
where $\mu_{0}=\unit[4\pi\times10^{-7}]{H/m}$ is the vacuum permeability
and $\omega=2\pi f$ is the circular AC frequency. For the AC frequency
$f=\unit[400]{Hz}$ used in this experiment we had $\delta\approx\unit[15]{mm}.$
This value is close to the optimal one which produces maximal flow
velocity at fixed current amplitude in this setup \cite{Moffatt-91}.

Typical melt velocity of $V=\unit[50]{mm/s}$ corresponds to the Reynolds
number $\textit{Re}=DV/\nu\approx7400,$ which implies turbulent flow
regime. Magnitude of the electromagnetic force is characterized by
the dimensionless interaction parameter $N=(DI_{0}\mu_{0}/\nu)^{2}\omega\sigma/\rho$
$(\approx3.9\times10^{9}$ for $I_{0}=\unit[300]{A}).$

           

\section{\label{sec:mtech}Measurement techniques}

The velocity distribution in the liquid metal was measured by two
methods: pulsed ultrasound Doppler velocimetry (UDV) and potential
probe with local magnetic field, which both are briefly described
below.

\subsection{Ultrasound Doppler velocimetry}

Ultrasound Doppler pulsed velocimetry (UDV) is a widely employed technique
for measuring velocities in liquid metal flows \cite{Takeda-87,Takeda-02,Cramer-etal-05}.
The method uses the phase shift of the ultrasonic wave scattered by
microscopic particles in the liquid to determine the distance to the
particles. By analysing the time correlation of the signals acquired
at different instants of time the device reconstructs the velocity
of particles along the ultrasonic beam. In this study we employed
DOP2000 Velocimeter model 2125 (`Signal Processing SA', Switzerland)
to measure the axial velocity profiles along the height of the container
in the flow of GaInSn driven by a two-phase AC magnetic field. Measuring
velocities of a few centimetres per second over a relatively small
height of the container required the highest available ultrasound
frequency of $\unit[10]{MHz}.$ For the pulse repetition frequency
of $\unit[1428]{Hz}$ used in the experiment and the speed of sound
in the liquid metal of $\unit[2740]{m/s},$ the velocity was measured
with the absolute error not exceeding $\unit[1]{mm/s}.$

The ultrasound wave was introduced into the liquid metal directly
through its free surface. A special care was taken to ensure a good
wetting of the UDV transducer by the liquid metal, which was essential
for the acoustic coupling. The vertical velocity profiles were measured
at the following 9 radial positions: $r=0,\pm5,\pm10,\pm15,\unit[\pm21]{mm}.$
The radial size of the UDV probe precluded measurements at larger
$r,$ i.e., closer to the side wall. At each radial position, $300$
samples of axial velocity profiles were recorded during a $\unit[35]{s}$
interval with $\approx\unit[8.5]{Hz}$ sampling rate and a $\unit[0.3]{mm}$
spatial resolution in the axial direction. These data were used to
find both the time average of the velocity and its standard deviation,
i.e. the intensity of velocity pulsations, at each of $9\times170$
spatial measurement points over the vertical cross-section of the
container. Since the measurements at different $r$ were taken at
different times, it was not possible to follow the temporal evolution
of the flow over the whole cross-section.

\subsection{Potential probe}

\begin{figure}
\begin{centering}
\includegraphics[height=0.25\textheight]{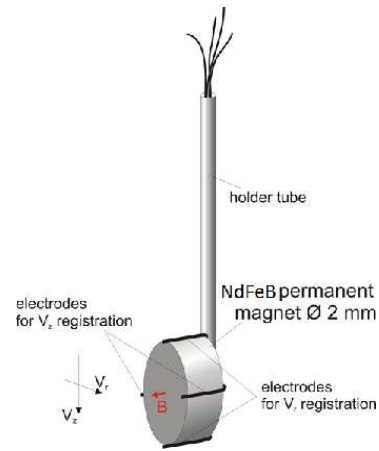} 
\par\end{centering}

\caption{\label{fig:vives-probe}Potential probe consisting of a small permanent
magnet with two pairs of electrodes. The horizontal and vertical pairs
of electrodes measure the opposite, that is the vertical and horizontal,
respectively, velocity components. }
\end{figure}

Potential probe, also known as the Vives probe \cite{Vives,Cramer-etal-06}, is
another commonly used technique for local velocity measurements in
low-melting-point liquid metals. The probe shown in Fig. \ref{fig:vives-probe}
consists of a small permanent magnet with two pairs of small electrodes.
The liquid metal flow passing between the electrode pairs in the magnetic
field of the probe induces a difference of the electric potential
between the electrodes proportional to the velocity in the respective
direction. Besides the low cost, the main advantage of the potential
probe over a single-channel UDV is the possibility to measure simultaneously
two components of local velocity with a high temporal sampling rate.
However, there are also several disadvantages. First, very low voltages,
typically about $\unit[0.1\ldots0.2]{\mu V/mm/s},$ need to be measured.
Second, quite long measurement times are required even at moderate
spatial resolutions to acquire the velocity distribution over the
whole cross-section of the container. Third, the temperature of the
liquid metal has to be kept sufficiently uniform to avoid thermo-electrically
induced potential perturbations. In the present experiment, the potential
probe was used only for a few measurements to check the UDV results.

\begin{figure}
\begin{centering}
\includegraphics[width=1\columnwidth]{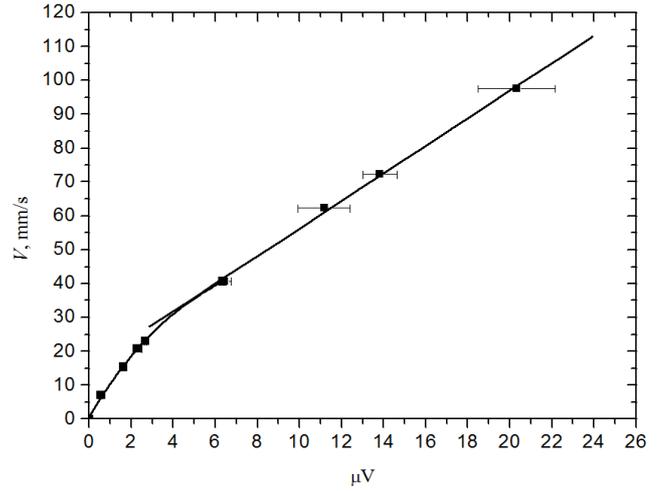} 
\par\end{centering}

\caption{\label{fig:calibr}Calibration curve for the potential probe with
horizontal bars showing the standard deviation of the signal induced
by the oscillations of the flow around the probe in the calibration
channel.}
\end{figure}

The probe used in this experiment to measure the axial and radial
velocity components was made of a cylindrical NdFeB permanent magnet
of $\unit[2]{mm}$ in both diameter and height, which was hold by
a stainless steel tube of $\unit[1]{mm}$ in diameter. The probe was
calibrated using an annular channel filled with GaInSn, which was
rotated about its centre with a well-defined angular velocity. Since
the sensitivity of the probe is nonlinear at low velocities (see Fig.
\ref{fig:calibr}), calibration was carried out for the whole range
of the expected velocities. The calibration data were fitted with
two mathematical expressions: a `sigmoid law' for low velocities and
a linear one for higher velocities. These expressions were further
used in the acquisition software to convert the voltage measurements
into velocities.

Measurements were taken at $121$ locations in the vertical cross-section
of the container. At each locationm the axial and radial velocity
components were recorded during a period of $\unit[100]{s}$ with
a $\unit[50]{Hz}$ sampling rate using a high-resolution 24-bit `Data
Translation' DT9821 acquisition board. The internal pre-amplifier
of the board with the gain of $64$ ensured a $\unit[20]{nV}$ `noise-free'
resolution in voltage at about the same RMS value of the noise. Besides
the instrumental noise, the accuracy of measurements was affected
also by thermally induced zero drift of the acquisition board. This
effect was minimized by keeping the acquisition board in a thermo-box,
which limited the total uncertainty of the electric potential measurement
to $\unit[450]{nV}$. For the sensitivity of the probe in the linear
regime $\unit[0.2]{\mu V/mm/s}$ (Fig. \ref{fig:calibr}), the uncertainty
of the velocity measurements was about $\unit[2.3]{mm/s}.$

\section{\label{sec:expres}Experimental results}

\begin{figure*}
\begin{centering}
\includegraphics[bb=30bp 70bp 598bp 630bp,clip,height=0.3\textheight]{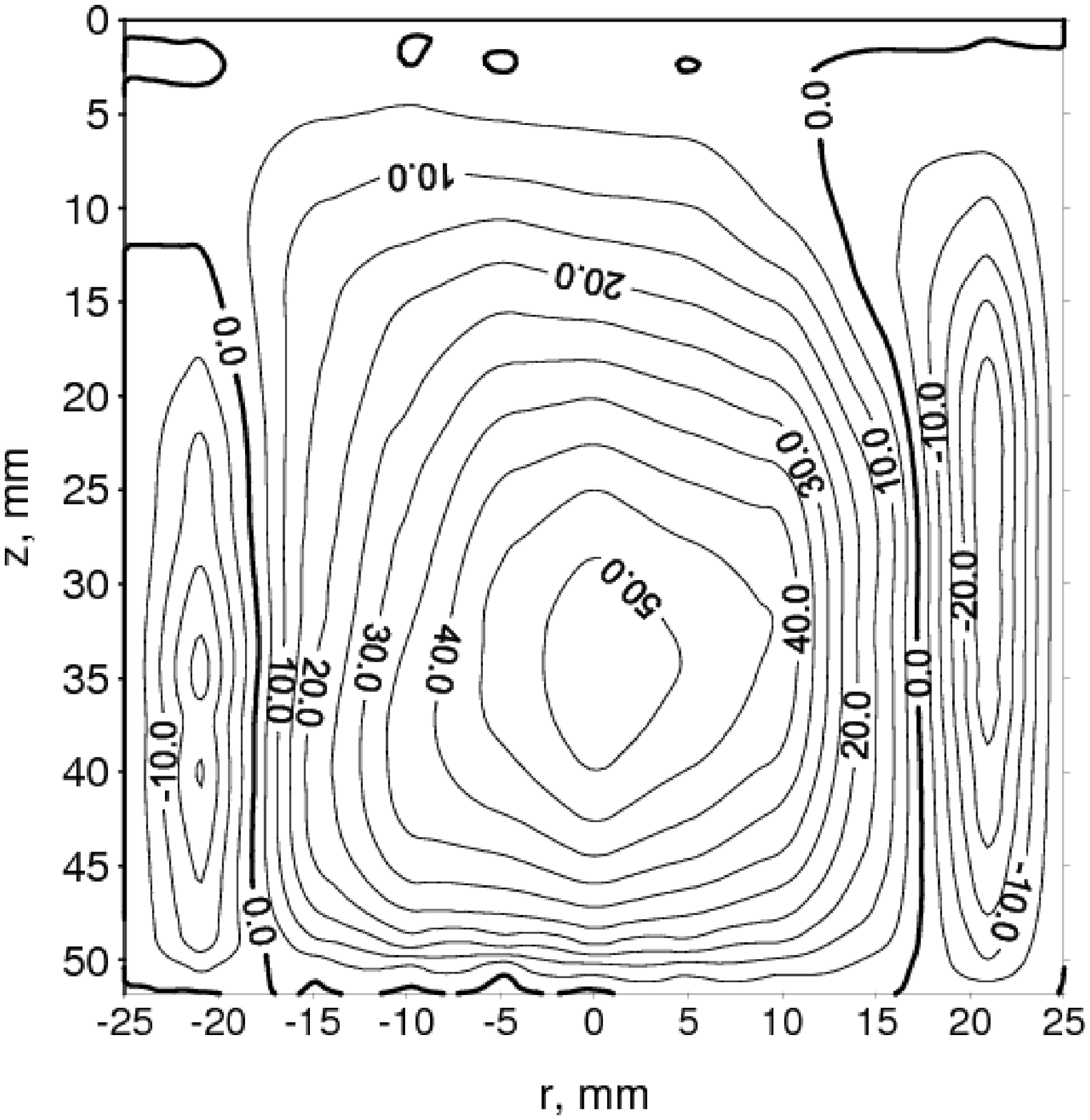}\includegraphics[bb=0bp 85bp 598bp 645bp,clip,height=0.3\textheight]{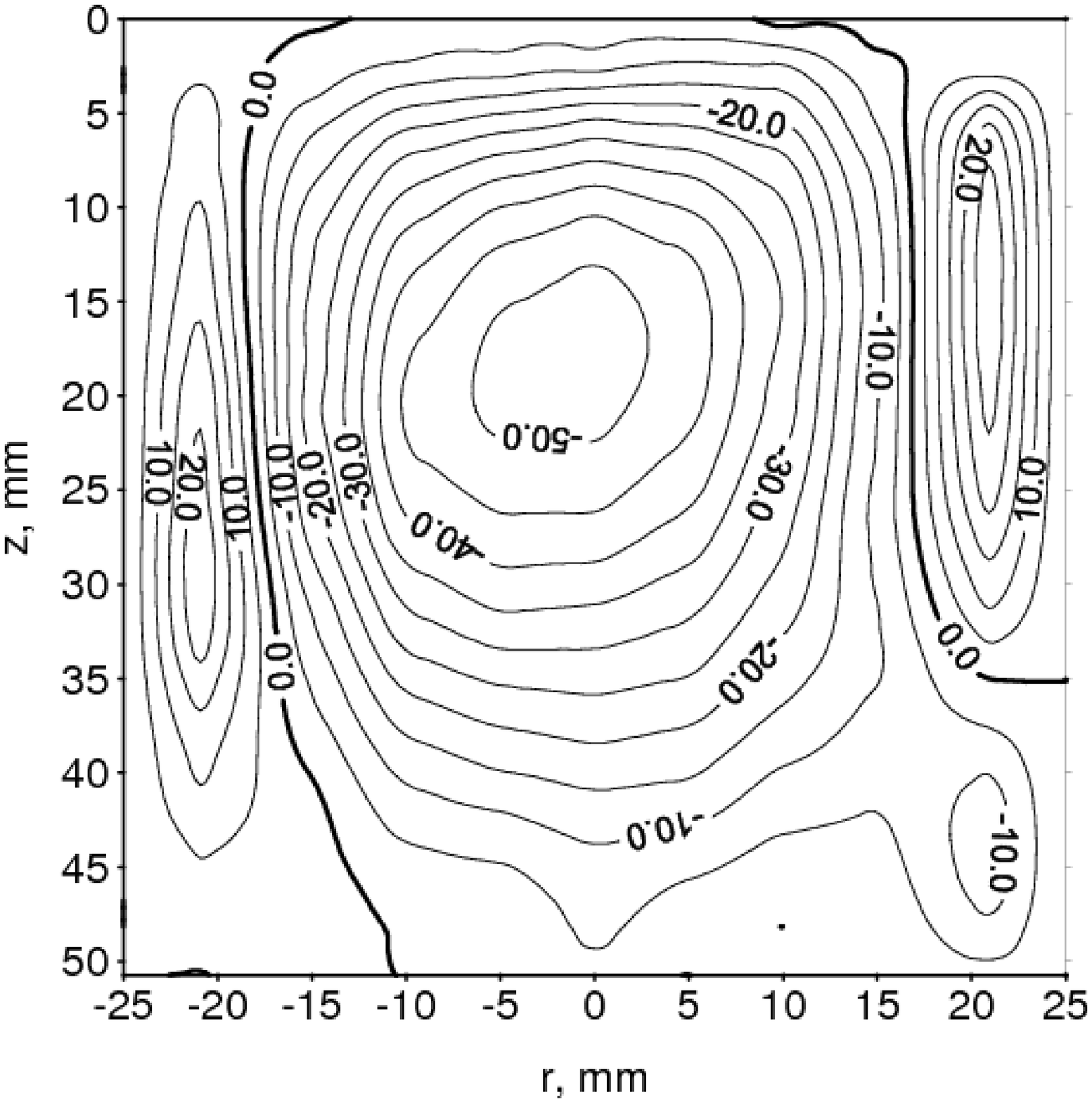}\\
(a)\hspace{0.5\columnwidth}(b) 
\par\end{centering}

\caption{\label{fig:UDV-I350-f90}Isolines of the axial velocity component
for the magnetic field travelling downwards (a) and upwards (b) which
correspond to the current in the upper coil lagging and leading, respectively,
in phase by $90^{\circ}$ relative to the current in the bottom coil.
In both coils the effective current is the same and equal to $I=\unit[350]{A}.$ }
\end{figure*}

\subsection{UDV measurements}

\begin{figure}
\begin{centering}
\includegraphics[width=1\columnwidth]{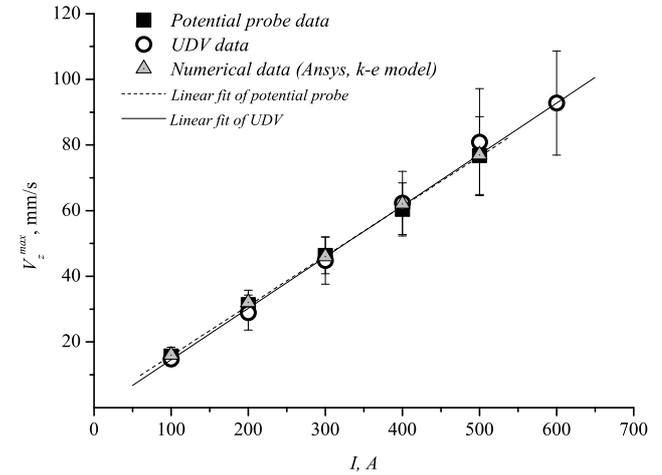} 
\par\end{centering}

\caption{\label{fig:Vmax-I-UDV+NUM}Maximal axial velocity measured by the
UDV technique (circles) and the potential probe (squares) along with
numerical results (triangles) versus the effective current $I$ at
the phase shift fixed to $\Delta\varphi=90^{\circ}.$ The error bars
show the standard deviation of turbulent velocity pulsations. }
\end{figure}

The basic effect of the phase shift between the currents in the top
and bottom coils on the flow structure is illustrated by Fig. \ref{fig:UDV-I350-f90}
which shows the isolines of the vertical velocity component for the
magnetic field travelling upwards (a) and downwards (b). Note that
the velocities are recorded as positive when the flow is directed
away, that is downward from the UDV probe which is placed at the top
of the liquid metal. Also, note that the velocity could be measured
by the UDV only up to $|r|\le\unit[21]{mm}$ and is then extrapolated
to the region near the side wall where it is supposed to obey the
no-slip condition. In both cases, the currents in the top and bottom
coils alternate with the same frequency of $f=\unit[400]{Hz}$ and
have the same effective values equal to $I=\unit[350]{A}.$ The only
difference is that in the first case, the current in the upper coil
lags in phase by $90^{\circ}$ while in the second case, it leads
by the same amount relative to the current in the bottom coil. As
the lagging current follows the leading one, the magnetic field effectively
travels from the former to the latter. And as it does so, it drags
the liquid metal near the side wall with it. Respectively, the magnetic
field travelling upward gives rise to regions of negative axial velocity
at the side walls, while there is a larger region of positive velocity
in the central part of the container where a downward axial jet forms.
The highest axial velocity is attained in the jet at the symmetry
axis ($r=0$) somewhat downstream from the mid-height of the container.
Switching the sign of the phase shift and thus the vertical direction
in which the magnetic field travels swaps both the direction of the
flow and its pattern about the mid-height of the container. The slight
axial asymmetry of the flow pattern observable in both cases may be
due to the magnetic field perturbations by the current connections.

\begin{figure*}
\begin{centering}
\includegraphics[height=0.3\textheight]{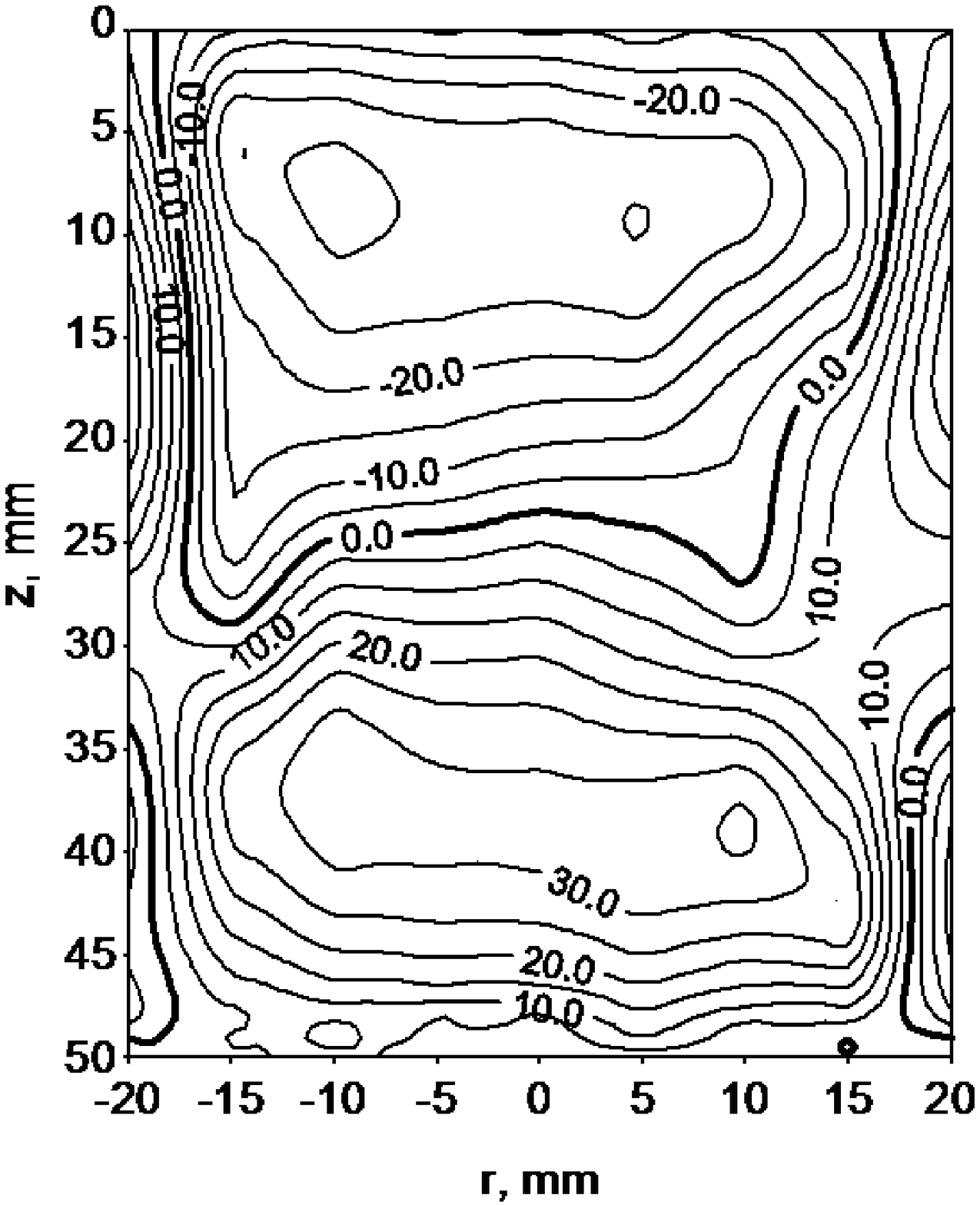}\put(-10,0){(a)}\quad{}\includegraphics[height=0.3\textheight]{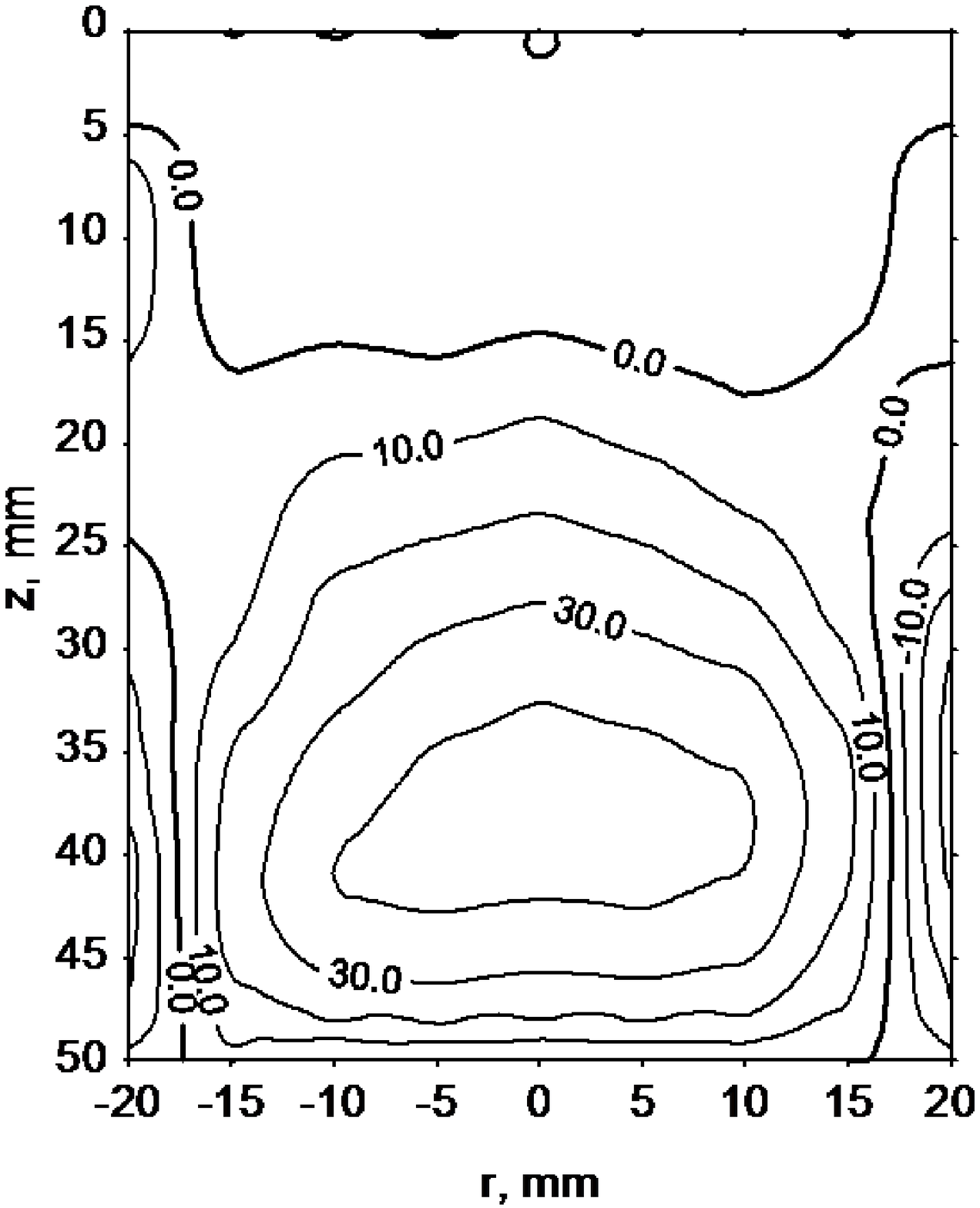}\put(-10,0){(b)} 
\par\end{centering}

\begin{centering}
\includegraphics[bb=0bp 0bp 407bp 500bp,clip,height=0.3\textheight]{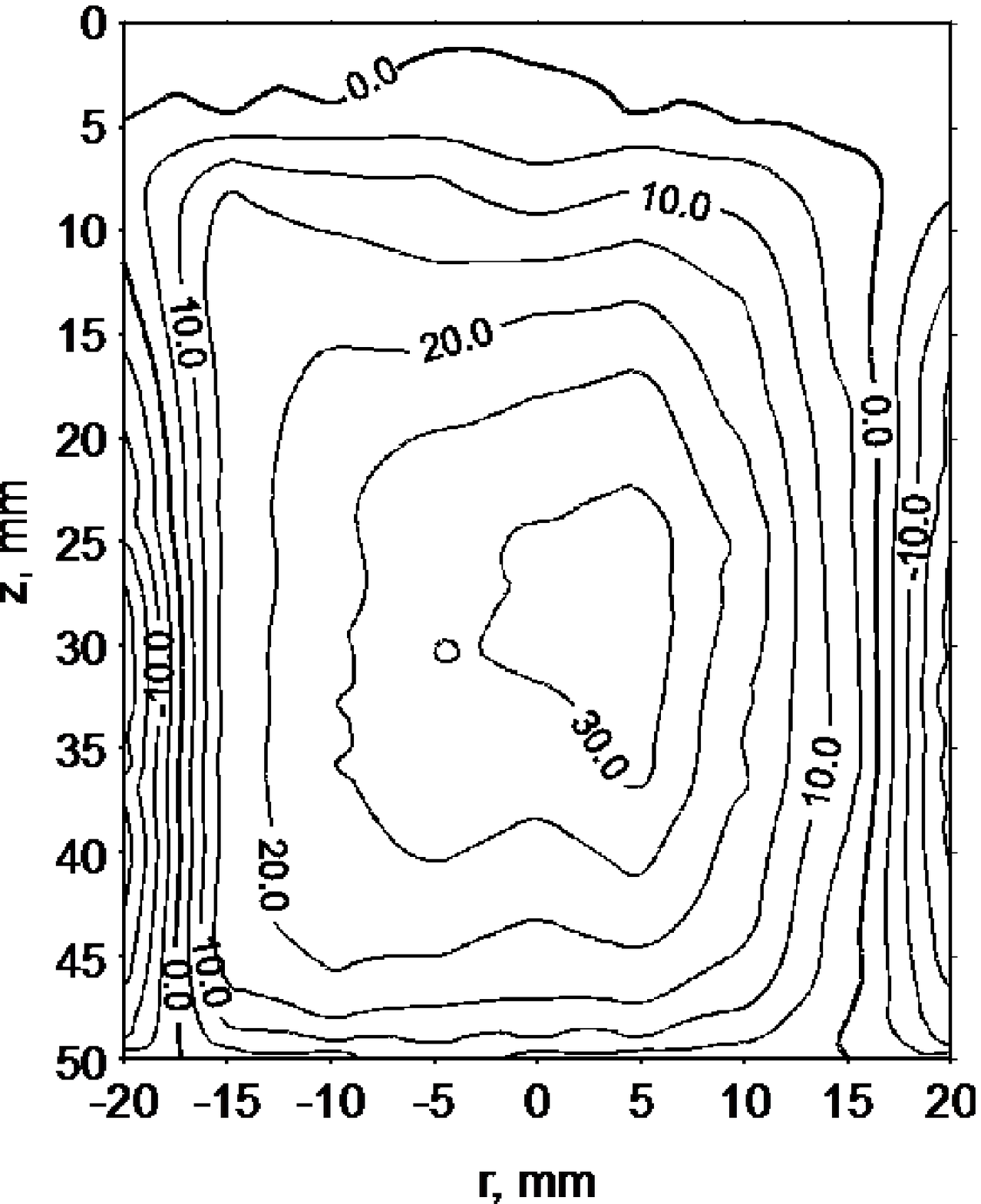}\put(-10,0){(c)}\quad{}\includegraphics[height=0.3\textheight]{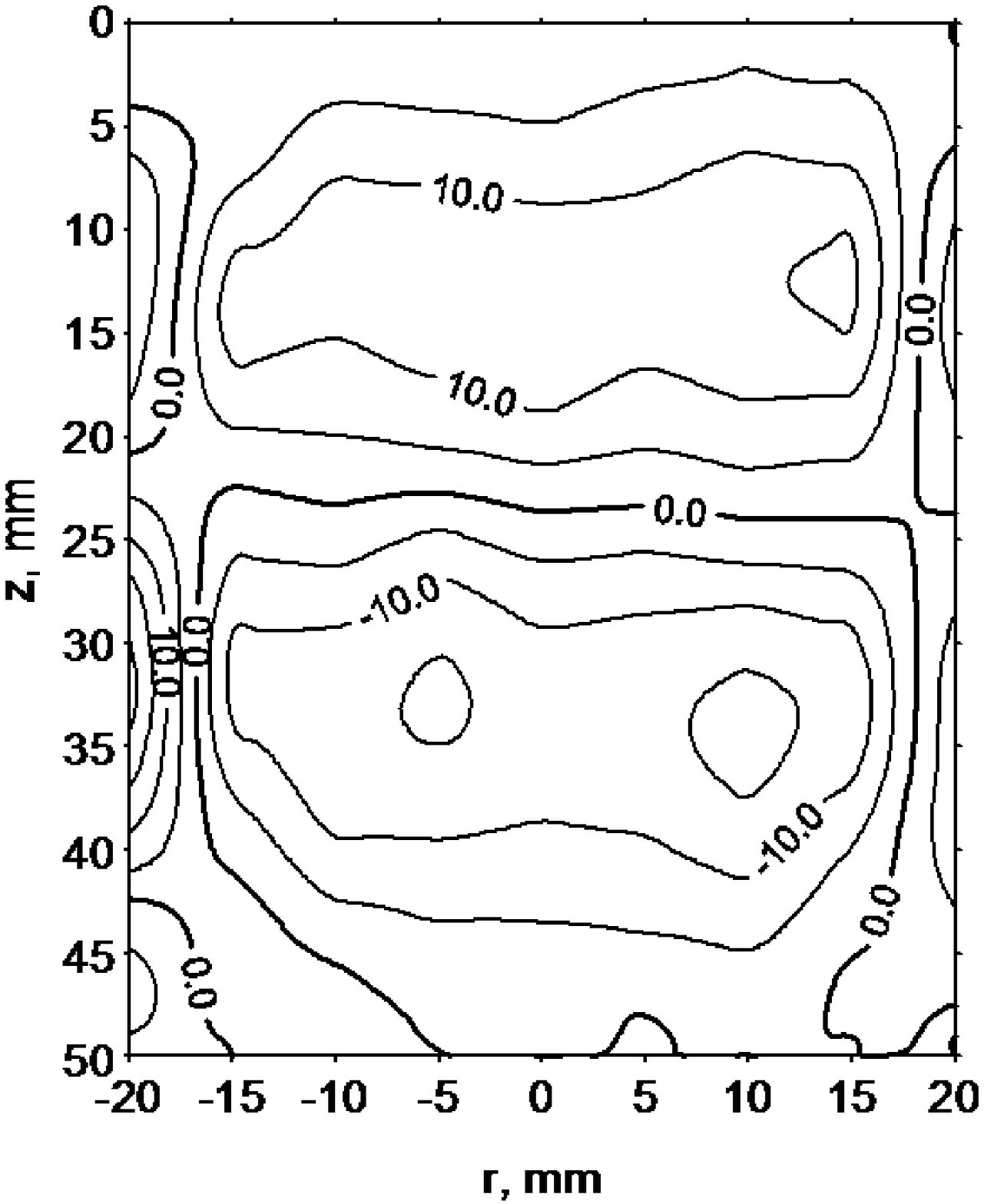}\put(-10,0){(d)} 
\par\end{centering}

\caption{\label{fig:Vz-phi}Isolines of the axial velocity component for various
phase shifts $\Delta\varphi$ between the currents in top and bottom
coils (a) $0^{\circ},$ (b) $30^{\circ},$ (c) $120^{\circ},$ (d)
$180^{\circ}.$ The case of $\Delta\varphi=90^{\circ}$ can be seen
in Fig. \ref{fig:UDV-I350-f90}(a)}
\end{figure*}

Two series of measurements were performed using the UDV. In the first
series, the effective current $I$ in both coils was increased simultaneously
from $100$ to $\unit[600]{A}$ while the phase shift between them
was kept fixed at $\Delta\varphi=90^{\circ}.$ The maximal axial velocity,
which for this case is plotted in Fig. \ref{fig:Vmax-I-UDV+NUM},
is seen to increase with the current almost linearly from $15$ to
$\unit[92]{mm/s}.$ Such a linearity, which results from the balance
of electromagnetic and nonlinear inertial forces \cite{SneMof82},
is characteristic for turbulent AC-driven flows \cite{TabFau85}.
In laminar but strongly nonlinear flow regime, velocity is expected
to vary as $\sim I_{0}^{4/3}$ \cite{Fautrelle-81,Sneyd-1993}.

Along with the velocity grow also its fluctuations which characterize
the level of turbulence in the flow. The flow pattern remains basically
similar to that shown in Fig. \ref{fig:UDV-I350-f90} (left panel)
with a slight axial asymmetry.

\begin{figure*}
\begin{centering}
\includegraphics[height=0.25\textheight]{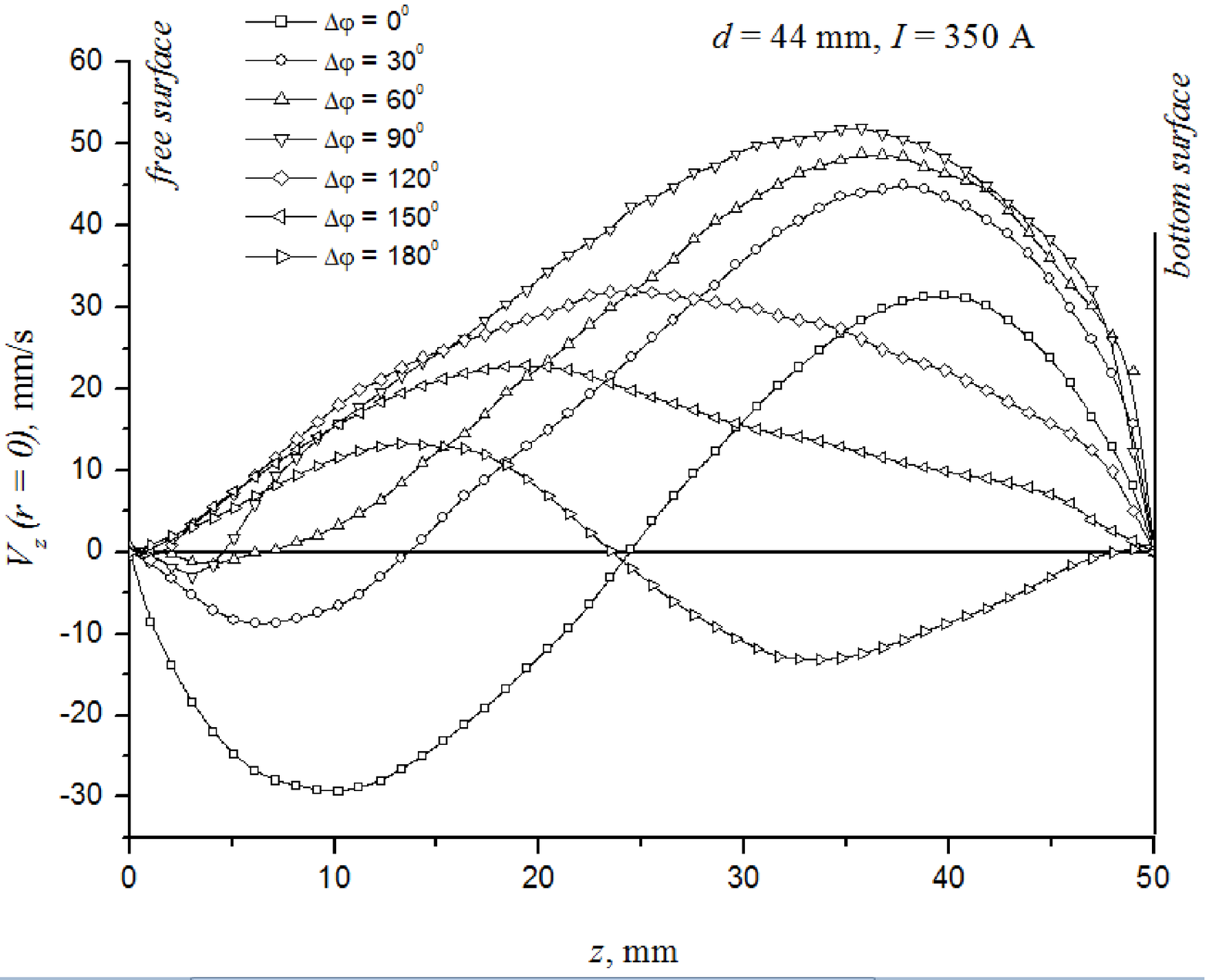}\includegraphics[height=0.25\textheight]{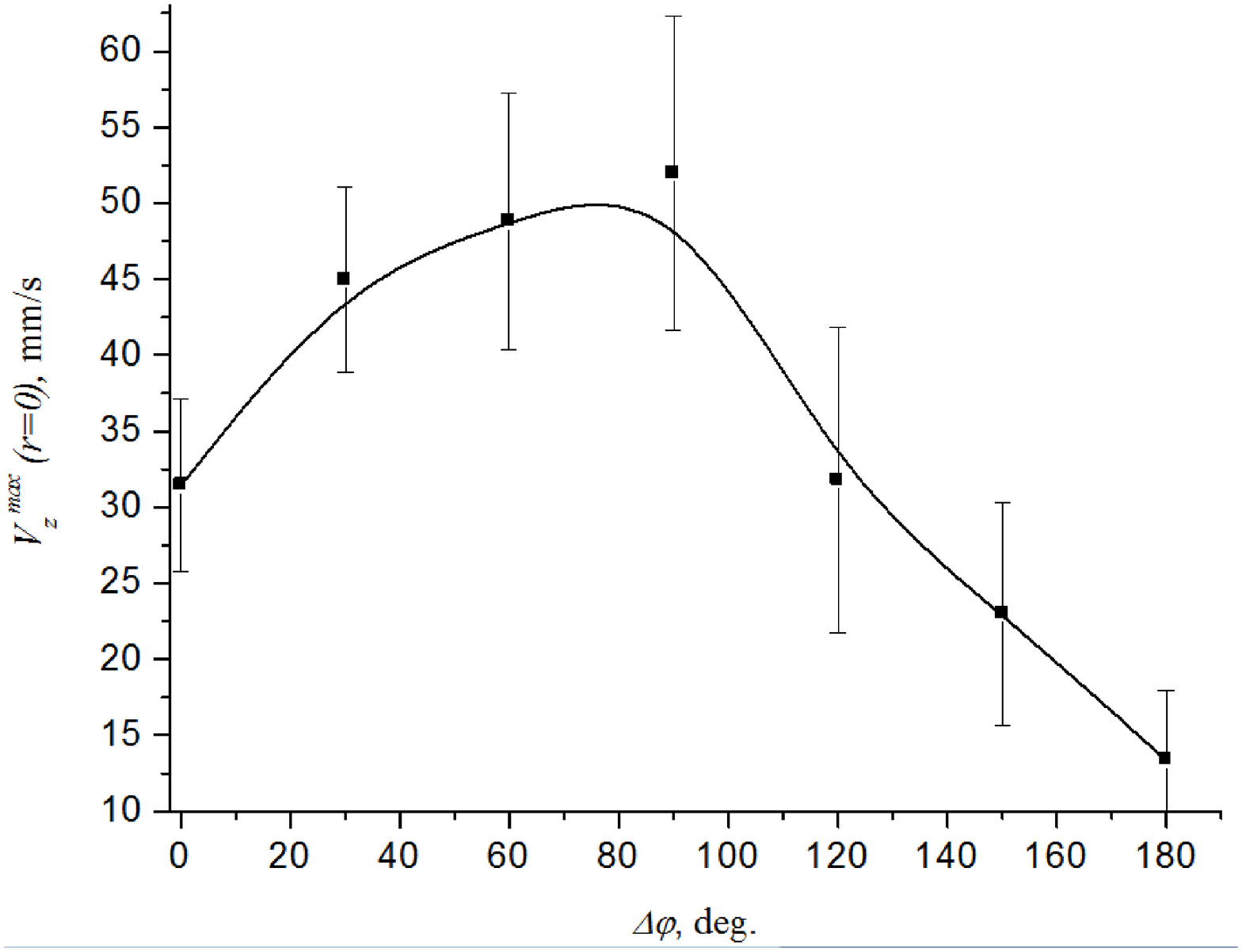}\\
(a)\hspace{0.5\columnwidth}(b) 
\par\end{centering}

\caption{\label{fig:Vz-r0}Axial velocity profiles along the symmetry axis
$(r=0)$ at different phase shifts between the currents in both coils
(a). The highest axial velocity along the symmetry axis and the RMS
of the associated turbulent velocity fluctuations versus the phase
shift between the currents (b).}
\end{figure*}

In the second series, the effective currents in both coils were kept
fixed at $I=\unit[350]{A}$ while the phase shift $\Delta\varphi$
between them was increased from $0^{\circ}$ to $180^{\circ}.$ Figure
\ref{fig:Vz-phi} shows the isolines of the axial velocity component
measured at four different phase shifts $\Delta\varphi$ in addition
to that of $\Delta\varphi=90^{\circ},$ which is plotted in Fig. \ref{fig:UDV-I350-f90}a.
The first case with $\Delta\varphi=0^{\circ}$ shown in Fig. \ref{fig:Vz-phi}a
corresponds to a standard single-phase AC magnetic field. This field
gives rise to a nearly compressing electromagnetic force with a maximum
at the mid-height between the coils, where it drives the liquid metal
radially inwards. As a result the typical two-vortex structure, which
can be inferred from Fig. \ref{fig:Vz-phi}a, arises. As seen in Fig.
\ref{fig:Vz-phi}b, a phase shift of $\Delta\varphi=30^{\circ}$ is
sufficient to break this two-vortex balance in favour of the bottom
one which is enhanced by the arising upward axial force while the
top vortex is suppressed by it. The axial flow keeps increasing with
the phase shift up to $\Delta\varphi=90^{\circ}$ where according
to Fig. \ref{fig:Vz-r0} it attains a maximum and then starts to decrease
when the phase shift is increased further. The lowest axial velocity
is attained at $\Delta\varphi=180^{\circ}$ which again corresponds
to a single-phase AC magnetic field. In contrast to the previous case,
the currents in both coils now flowing in the opposite directions
tend to cancel out each other's magnetic field. Thus, the electromagnetic
pinching force cancels out in the mid-plane between the coils where
the liquid is now pushed by the pressure gradient radially outwards.
This gives rise to a two-vortex structure shown in Fig. \ref{fig:Vz-phi}d,
which is similar to that for $\Delta\varphi=0^{\circ}$ seen in Fig.
\ref{fig:Vz-phi}a except for the opposite flow direction and a twice
lower velocity.

\subsection{Potential probe measurements }

\begin{figure*}
\begin{centering}
\includegraphics[bb=0bp 0bp 744bp 724bp,height=0.3\textheight]{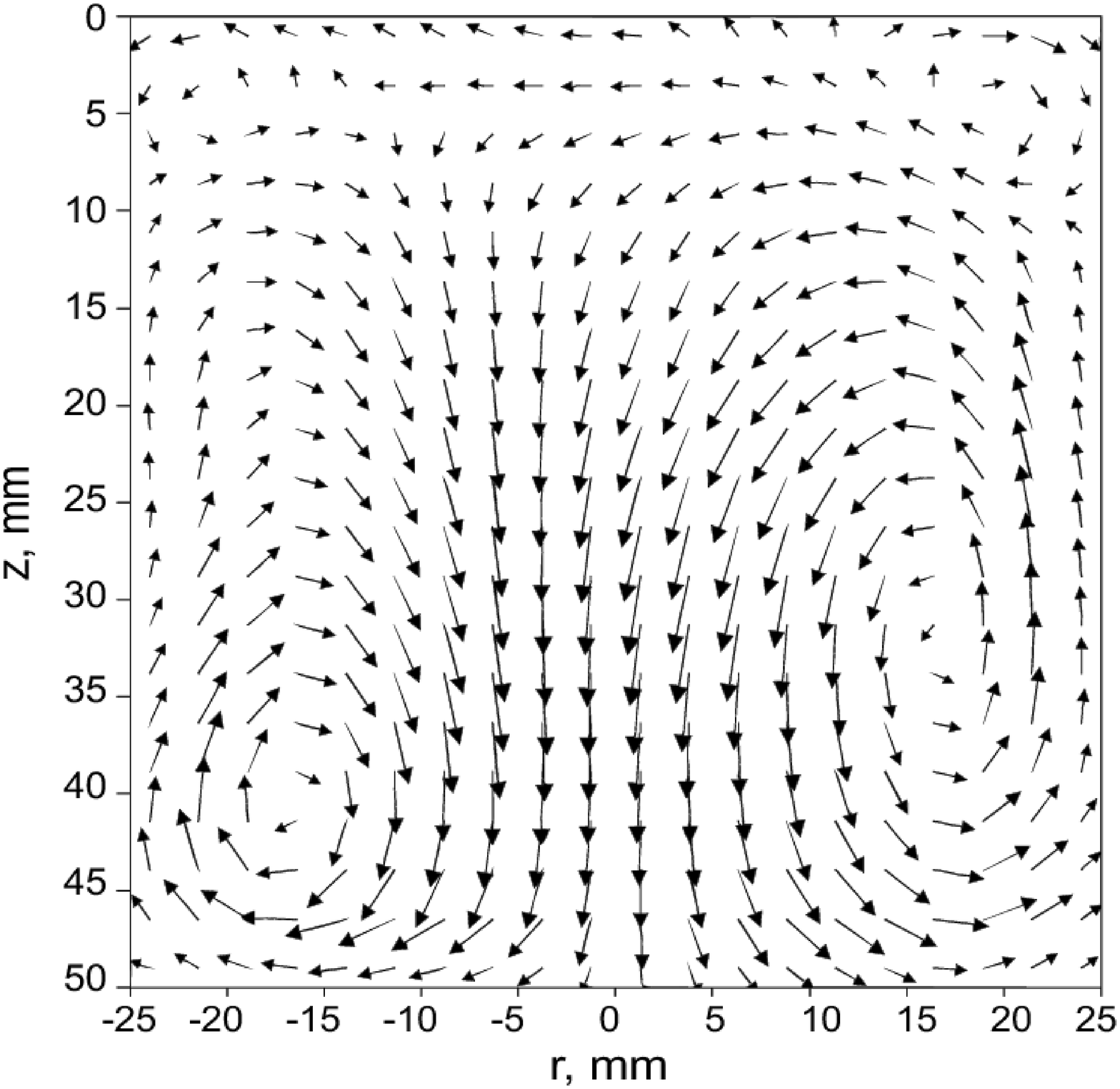}\includegraphics[bb=0bp 265bp 780bp 1000bp,clip,height=0.3\textheight]{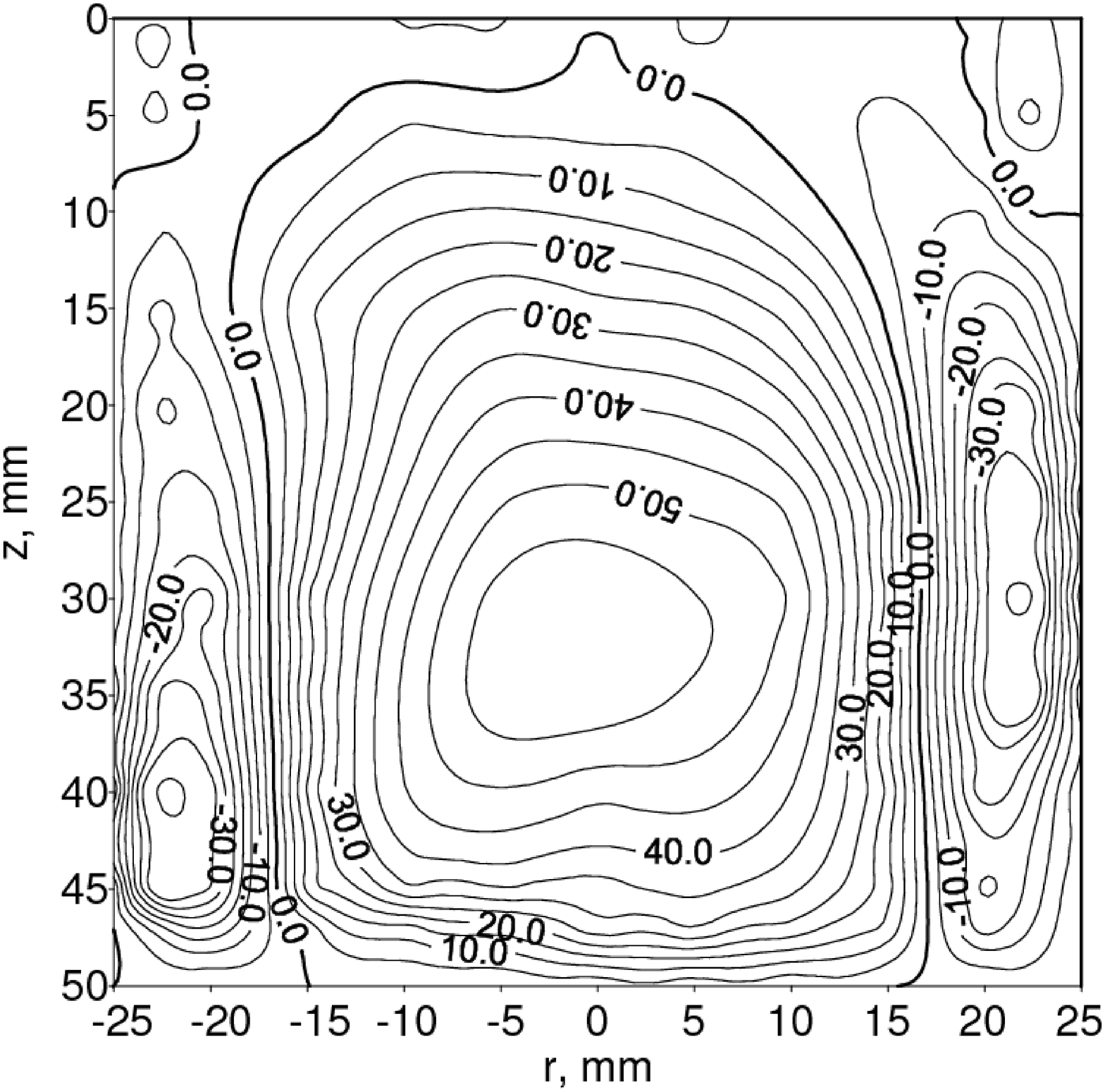}\\
(a)\hspace{0.5\columnwidth}(b) 
\par\end{centering}

\caption{\label{fig:POT-I350-f90}The velocity field in the vertical cross-section
(a) and the isolines of the axial velocity component (b) measured
by the potential probe at $I=\unit[350]{A}$ and $\Delta\varphi=90^{\circ}.$ }
\end{figure*}

Potential probe was used only in one case for $I=\unit[350]{A}$ and
$\Delta\varphi=90^{\circ},$ which was then compared with the UDV
measurements and numerical results. The measured velocity field and
the isolines of axial velocity component are shown in Fig. \ref{fig:POT-I350-f90}.
The latter is seen to agree well with the corresponding UDV measurements
shown in Fig. \ref{fig:UDV-I350-f90}a. Even the slight axial asymmetry
is reproduced. This probe could reach the radial positions up to $r=\pm\unit[23]{mm}$
so approaching the side wall closer than the UDV well into the region
of the ascending flow. The full flow pattern shown in Fig. \ref{fig:POT-I350-f90}
includes an extrapolation of the velocity distribution up to the side
walls with no-slip boundary condition applied. As seen in Fig. \ref{fig:Vmax-I-UDV+NUM},
there is also a good agreement between the maximal values of axial
velocities measured by the potential and UDV probes.

\section{\label{sec:comp}Comparison with numerical results}

The problem was also modelled numerically using the finite element
code ANSYS in axially symmetric approximation. First, the ANSYS Emag
solver was used to solve the electromagnetic part of the problem.
At this stage, the effect of liquid metal flow on the AC magnetic
field was neglected following the inductionless approximation which
is commonly used in the liquid metal magnetohydrodynamics. The cross-section
of metal was discretized using a rectangular uniform grid with $50\times100$
elements in radial and axial directions, respectively. The distribution
of the magnetic field was computed in both the liquid metal and the
surrounding areas for various coil currents and their phase shifts.
From those data, the distribution of the electromagnetic force in
the liquid metal was computed and then passed on as a body force to
the FLOTRAN hydrodynamic solver of ANSYS. Using this solver with a
standard $k$-$\varepsilon$ model of turbulence and no-slip boundary
conditions at the container walls, the flow of liquid metal driven
by AC magnetic field was computed using the same grid for the melt
region.

\begin{figure}
\begin{centering}
\includegraphics[width=1\columnwidth]{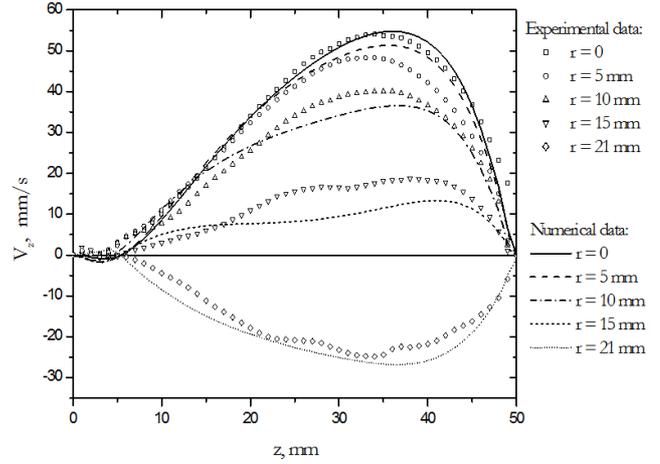} 
\par\end{centering}

\caption{\label{fig:Vzr-UDV-NUM}Measured (UDV) and computed profiles of the
mean axial velocity along the height of the container at various radii
for $I=\unit[350]{A}$ and $\Delta\varphi=90^{\circ}.$ Maximal RMS
value of turbulent velocity pulsations is shown in Fig. \ref{fig:Vz-r0}(b). }
\end{figure}

First, as seen in Fig. \ref{fig:Vmax-I-UDV+NUM}, the numerical results
demonstrate a good agreement for the maximal values of the axial velocity
measured by the UDV technique at various currents and a phase shift
fixed to $\Delta\varphi=90^{\circ}$. There is also a reasonable agreement
between the measured and computed profiles of the axial velocity along
the height of the container at various radii, which are plotted for
$I=\unit[350]{A}$ and $\Delta\varphi=90^{\circ}$ in Fig. \ref{fig:Vzr-UDV-NUM}.
The slight axial asymmetry in the measured velocity distribution,
which may be due to the current connectors, was eliminated in the
comparison by taking the mean value between positive and negative
radii.

\section{\label{sec:sum}Conclusion}

The velocity distribution in the flow of GaInSn eutectic alloy driven
by a two-phase inductor in a cylindrical container was measured by
ultrasonic Doppler velocimetry and local potential probes. The flow
was also computed numerically, and a reasonable agreement with the
experimental results was found. It was demonstrated that the flow
structure can be modified from the typical two toroidal vortices into
a single vortex when the currents in the two coils are made to alternate
with a certain phase shift. The vortex is strongest for a given current
amplitude when the phase shift between the currents in both coils
is about $90^{\circ}.$ The obtained results may be useful for the
design of combined two-phase electromagnetic stirrers and induction
heaters for metal or semiconductor melts in applications where a well-defined
stirring is required.

\end{document}